# Nanoscale switching characteristics of nearly tetragonal $BiFeO_3$ thin films

Dipanjan Mazumdar[1], Vilas Shelke[1], Milko Iliev[2], Stephen Jesse[3], Amit Kumar[3], Sergei Kalinin[3], Arthur Baddorf[3], *and* Arunava Gupta[1]

1. Center for Materials for Information Technology, University of Alabama, Tuscaloosa, AL 35487 USA

2. Texas Center for Superconductivity, University of Houston, Houston, TX 77204-5002, USA

3. Center for Nanophase Material Sciences , Oak Ridge National Laboratory, Oak Ridge, TN 37831 USA

Abstract

We have investigated the nanoscale switching properties of strain-engineered $BiFeO_3$ thin films deposited on $LaAlO_3$ substrates using a combination of scanning probe techniques. Polarized Raman spectral analysis indicate that the nearly-tetragonal films have monoclinic (*Cc*) rather than *P4mm* tetragonal symmetry. Through local switching-spectroscopy measurements and piezoresponse force microscopy we provide clear evidence of ferroelectric switching of the tetragonal phase but the polarization direction, and therefore its switching, deviates strongly from the expected (001) tetragonal axis. We also demonstrate a large and reversible, electrically-driven structural phase transition from the tetragonal to the rhombohedral polymorph in this material which is promising for a plethora of applications.

Since the 2003 report of high polarization in $BiFeO_3$ (BFO) thin films, research activity in multiferroic materials have generated interest across many scientific disciplines [1, 2]. As a single phase material, BFO has shown a wealth of properties, which includes simultaneous ferroelectric and magnetic order, and photo-voltaic effect [3]. Additional functionalities are expected when used in hetero-structures exploiting the magneto-electric (ME) coupling, or as a barrier material in a tunnel junction configuration with or without magnetic electrodes [4-11].

From a purely materials point-of-view BFO possess significant challenges. In its bulk form BFO assumes a low-symmetry rhombohedral (*R3c* symmetry) structure with polarization along the (111) direction, implying that the (001) films have both an in-plane and out-of-plane polarization components and a total of eight possible domains are possible in the film[12]. As a consequence, switching is often incomplete, as evidenced through incomplete hysteresis loops. To reduce the number of possible switching variants one route has been through the use of high-miscut vicinal substrates [13-15]. This has also been seen to reduce the coercive field, a critical parameter when it comes to device engineering [15].

But more promising avenues of BFO domain engineering has opened up very recently after a few groups reported successful fabrication of the tetragonal (or nearly tetragonal to be precise) variant of BFO [16-18]. The remarkable structural characteristic of this phase is its extremely large c/a ratio of around 1.25 achieved through the substrate strain effect (or strain engineering) on $LaAlO_3$ (LAO) and $YAlO_3$ (YAO) substrates [16, 17]. The high in-plane compressive strain (BFO has a lattice mismatch of -4.5% and -6.8% on LAO and YAO substrates, respectively) results in a large expansion of the out-of-plane lattice parameter, much higher than expected based on normal elastic energy considerations of rhombohedral bulk BFO. Therefore, this nearly tetragonal polymorph has been speculated to be a distinctly different phase and not a strained variant of the rhombohedral phase [17]. Epitaxial stabilization of the tetragonal phase is of significant interest as it has been theoretically predicted to have a high polarization value of over 150 $\mu C/cm^2$, which is about 50% higher than its rhombohedral counterpart [18-20]. Additionally, the switching process can, in principle, be simpler and more efficient compared to the rhombohedral phase, making it further desirable.

Films grown on LAO susbtrates are in particular interesting because they show a mixed tetragonal (T) and rhombohedral (R) character with the relative fraction dependent on the film thickness [16]. As the strain in the film is gradually released the rhombohedral phase content increases, due to its overwhelming lower ground state energy as compared to the tetragonal phase[16, 19]. Such a mixed phase is well known in other ferroelectric systems. Common piezoelectric materials, such as PZT ($PbZrO_3$-$PbTiO_3$), PMN-PT ($PbMg_{2/3}Nb_{1/3}O_3$-$PbTiO_3$), display a structural transition in their phase diagram as a function of the composition [21-23]. Such a boundary, termed as the morphotropic phase boundary (MPB), separates regions of tetragonal and rhombohedral symmetry. Further these systems show enhanced electromechanical properties due to the existence of such MPBs. However, a single-phase lead-free material with similar properties is more desirable from present day application point of view. Therefore, the possibility of observing such a behavior is an enticing choice in the mixed tetragonal-rhombohedral BFO grown on LAO. From now on we shall denote the nearly tetragonal phase with “T” and the rhombohedral phase with “R”.

Epitaxial BFO thin films have been grown in the thickness range of 70-200 nm using the pulsed laser deposition (PLD) technique on (001)-oriented LAO substrates, with and without bottom $SrRuO_3$ (SRO) electrodes of 40 nm thickness. Details of the sample preparation technique have been reported elsewhere[24]. In Fig. 1 we show the results of our X-ray analysis. Detailed θ-2θ x-ray diffraction measurements have been performed to ascertain the structural phases and out-of -plane lattice constant of chemically single-phase BFO. In Fig. 1(a) we plot the θ-2θ scans in 35-50 degree range for the 70, 100 and 200nm samples deposited directly on LAO substrate and an 80 nm thick BFO film buffered with 40nm $SrRuO_3$ (SRO). Very strong (00*l*) peaks ((002) peak shown in Fig. 1(a)) are observed on films deposited directly on LAO substrate corresponding to an out-of-plane lattice parameter value of 4.65±0.02 Å. This c-parameter value agrees very well with the recent reports of Zeches *et al.* and Bea *et al.*[16, 17]. Also, the peak intensity is stronger for the 70 and 100 nm film compared to 200 nm film, which is consistent with decreasing T-phase content with thickness. The film with SRO bottom electrode, on the other hand, shows a diffraction peak close to the bulk rhombohedral peak position of BFO (c~3.96 Å). Epitaxy analysis have been carried on the nearly-tetragonal BFO sample by performing phi-scans on its (101) reflection as shown in Fig. 1(b). Each peak of the phi scan show three sub-features pointing to the existence of three variants of the tetragonal phase (Inset of Fig. 1(b)). Asymmetric reciprocal-space maps (RSM) of the (103) and (013) tetragonal BFO reflection also show a three-variant structure, implying that the tetragonal structure is distorted with monoclinic tilts (Fig. 1(c), RSM of (013) reflection is now shown). Such distorted variants have been predicted to be lower in energy compared to the pure, undistorted tetragonal structure[17, 25]. From the peak position of the strongest spot we calculate an in-plane lattice constant of 3.73A and 3.75Åfor the 70 and 100 nm films, respectively, quite close to the pseudo-cubic lattice parameter of the LAO substrate (3.79Å). Therefore, we find our samples to be highly strained with a high c/a value of ~1.25 (see figure 1(d)), in excellent agreement with existing experimental and theoretical values[16-20]. However, the presence of in-plane oriented domains cannot be ruled as the strong LAO substrate peak very nearly coincides with the in-plane tetragonal lattice constants. Also, conclusive signature for the rhombohedral structure has not be picked up in these samples from x-ray analysis. But all BFO films grown with buffered SRO bottom electrodes are almost completely relaxed with a c-value of 3.97±0.01 Å, and insensitive to the film thickness studied here (80-600 nm). The presence of SRO buffer layer favors faster strain relaxation through formation of misfit dislocations[26].

Even more informative with respect to the type of structure are the polarized Raman spectroscopy data. Fig.2 shows the polarized Raman spectra of the films of Fig. 1a obtained under a microscope (probe spot 1-2 μm diameter) with *xx*, *xy*, *x'x'*, and *x'y'* scattering configurations using 488 nm excitation. The first and second letters in these notations denote, respectively, the polarization of incident and scattered radiation along $x\|[100]_c$, $y\|[010]_c$, $x'\|[110]_c$, or $y'\|[-110]_c$ quasicubic directions of the LAO substrate. The LAO contribution to the original spectra has been subtracted using the Spectral Subtract (GRAMS AI) software.

As is clear from Fig. 2, the tetragonal-like (70, 100, and 200 nm BFO/LAO) and rhombohedral (80 nm BFO/40 nm SRO/LAO) films exhibit significantly different spectra. The fact that the spectra in the *xx-x'x'* and *xy-xy* pair are different provides clear evidence that all films are epitaxial. On the other hand, the identity of the *xx-yy* and *x'x'-y'y'* spectra for the tetragonal phase is consistent with $(001)_t$ tetragonal plane being parallel to the surface. A careful study of numerous points on the 70-200 nm BFO/LAO films show that tetragonal-like and rhombohedral domains of micrometer size coexist on the film surface with the rhombohedral phase gradually becoming the dominant phase with increasing thickness. As to the spectra of the rhombohedral phase, they are identical to those previously reported[27,28] and their polarization properties can be explained accounting for the coexistence in the scattering volume of twin variants with four different orientation of the rhombohedral axes. Raman study of the tetragonal-like phase has not been reported before and the observation of more than one mode of $B_1$-type symmetry indicates a more complicated structure. On careful analysis based on symmetry and the number of Raman peaks, we find strong evidence that the nearly-tetragonal structure has monoclinic *Cc* symmetry instead of simple *P4mm* tetragonal symmetry. A detail analysis of the Raman spectra of tetragonal BFO is provided elewhere.[27] We will only mention here that for the currently accepted tetragonal *P4mm* BFO structure one expects in total eight ($3A_1 + B_1 + 4E$) Raman modes. Only $A_1$ and $B_1$ modes can be observed from the (001) surface. The $A_1$ modes are allowed in all parallel and forbidden in all crossed polarizations, the $B_1$ mode is allowed with parallel polarizations along $[100]_t$ and $[010]_t$ and crossed $[110]_t[-110]_t$ polarization configuration, but forbidden with parallel $[110]_t$ $[110]_t$ (or $[-110]_t[-100]_t$) and crossed $[110]_t[-110]_t$ configurations. The only configuration where neither $A_1$, nor $B_1$ modes are allowed is $[100]_t[010]_t$. Unlike $x'y'$, no strong spectral structure are observed in the $xy$ spectra of the tetragonal phase, which strongly suggest that the tetragonal $[100]_t$ and $[010]_t$ directions are parallel to the quasicubic $[100]_c$ and $[010]_c$ directions of the LAO substrate.

Ferroelectric (FE) domains of our samples were investigated using the piezoresponse force microscopy technique (PFM) performed on a commercial atomic force microscope (Asylum Research, Cypher Model). Nanoscale FE switching properties were investigated using the newly-developed switching spectroscopy PFM (SS-PFM) technique [28, 29] as implemented in the Cypher AFM. Vertical PFM measurements were performed near the contact resonance of the tip-sample configuration in order to obtain high signal-to- noise ratio through resonance-enhancement. In addition, both vertical (out-of-plane) and lateral (in-plane) PFM images were performed far from resonance. Pt-coated conductive tips (Olympus AC240TM) were used with the AC-bias drive amplitude maintained typically between 1-2V. SS-PFM measurements were performed in a bandwidth of 50 kHz around contact resonance.

In Fig. 3(a) we show the topography and near-resonance vertical PFM amplitude and phase data of the 80 nm BFO film on buffered SRO. Mosaic domain patterns (Fig. 3a(ii) and 3a(iii)) are observed with little correlation with topography, and verified by scanning different areas of the sample. Vertical and lateral PFM images taken at frequencies far from resonance (see supplementary figure S1) further confirm the irregular domain configuration. Interestingly, the lateral PFM data shows a higher contrast than the vertical data, implying that the domains can have strong in-plane orientation. These observed patterns are reminiscent of rhombohedral BFO films deposited on SRO-buffered STO substrates [30, 31], and consistent with our Raman spectra and X-ray diffraction analysis. However, this is unlike the results obtained with $La_{1-x}Sr_xMnO_3$ (LSMO) or $La_{1-x}S_xCoO_3$ (LSCO) buffer layer which have been shown to preserve the substrate strain [16, 17] and tetragonality.

Topography and ferroelectric domains (near-resonance vertical PFM) of the 100 and 200 nm tetragonal-BFO films are shown in fig. 3(b) and (c). The surface morphology of BFO deposited directly on LAO consists of two distinct features (Fig. 3b(i) and 3c(i)). The majority of the surface shows extremely flat morphology interjected with one- dimensional arrays of 250-350 nm long rod-shaped trenches aligned side by side and extending to well over microns in length. These arrays, in many instances, run approximately perpendicular to each other but in general have no fixed orientation with respect to the substrate edges. In addition, uniformly dispersed nano-island depressions are observed. These trenches have been identified as being consistent with the rhombohedral phase[16] and can be thought of as due to their low c/a ratio compared to the nearly-tetragonal phase as depicted in the schematic figure 1(d). Interestingly, the flat, nearly-tetragonal surface show a step-and-terrace morphology in all the films investigated. A straight-line sectional analysis gives a step-height value of 4.5±0.8Å which is very close to the lattice constants obtained from X-ray diffraction and RSM analysis (fig.1).

A weak but distinct PFM contrast is observed in the tetragonal areas with a striped-domain pattern oriented 45 degrees to the substrate [100] and [010] directions, i.e, along the [110] and [-110] direction (fig 3b(ii)&(iii), 3c(ii)&(iii)) both in the 100 and 200 nm films. The stripes are approximately microns in length with a period of 70-80 nm and form flux-closed states with 90 degree domain walls as indicated by the black arrows in Fig. 3b(iii). The weak vertical contrast suggests that the domains might be predominantly in-plane. For further analysis, we have performed detailed vertical and lateral PFM measurements far from contact resonance (~30 KHz) as shown in the supplementary figures S2-S3. The vertical response shows uniform phase while the lateral response has strong phase contrast, strongly suggestive of in-plane oriented domains in the tetragonal phase. The vertical signal is remarkably enhanced (both for near- and off-resonance images) at the interface of the tetragonal and rhombohedral phase. Even though crosstalk interference cannot be ruled out completely given the large changes in topography, the enhancement is consistent with higher electro-mechanical response at the phase boundary. The inside of the rhombohedral trenches have weaker out-of-plane signal and stronger in-plane signal than the interface. Similar features are also observed in the PFM date obtained from the 200 nm thick film. The observation of in-plane oriented domains is quite contrary to the expectation from a nearly-tetragonal unit cell. In the work of Bea *et al.*[17], they have speculated the polarization direction to be in the (110) plane. Recent theoretical work of Hatt *et al.* [25] show that the angle between the polarization vector and the [001] direction to be extremely sensitive to small changes in the in-plane lattice parameter, especially for strains values encountered with LAO substrate, and that the polarization direction could be closer to (111) direction – similar to the rhombohedral phase.

We find that the nearly-tetragonal BFO films show reversible, electric-field driven phase transition between the R and the T phase. Large, hysteretic structural changes are observed when alternately a ± 20 V single voltage pulse of 1 sec duration is applied at the topographic center through the scanning probe tip in the sequence shown in fig 4(a)-(d). R-phase like trenches are formed on applying a sufficiently strong negative voltage pulse, which transforms back to the T-phase like topography upon reversing the polarity (Fig. 4(a)-(d)). On this particular film of 100 nm thickness, switching is observed only at voltages pulses above 15V and 250 millisecond pulse width. The R-phase like trench obtained in Fig. 4(b) and 4(d) are measured to be 2.3 nm deep, which is higher than the average trench-depth measured in the as-deposited configuration. Based on this observation, we speculate that the energy barrier separating the T and R phases is small and accessible by applying an electric field (figure 4(e) right). This is unlike the bulk form, where the R-phase is overwhelmingly the favored ground state and can be thought of as having a large energy barrier with the metastable bulk tetragonal phase, if it can be stabilized (figure 4(e) left). Therefore, observing phase transitions between these states is practically impossible. But such is not the case of highly-strained BFO films such as on LAO substrates

To investigate further, we performed additional topographic and PFM investigations to up to ±100 V. The absence of any bottom electrode allows us to apply large voltages without creating dielectric breakdown (Fig. 5). Voltage pulses are applied to the topographic center in the following sequence: -25V, -50V, -75V, -100V, +25V, +50V, +75V, +100V and PFM/AFM micrographs are taken after each pulse. In Fig. 5 we show images after the application of -25V (a), -75V (b), +25V (c), +75V (d). The as-is configuration of this area is the same as shown in Fig. 3(b). Consistent with the results of figure 4, application of negative voltages pulses induces local structural transition in the form of trenches. The density of such trenches increases steadily upon increasing the voltage to -100 V (figure 5(d)). The PFM images are qualitatively identical to the as-deposited states, thereby confirming that the trenches are indeed the rhombohedral phase. The tetragonal background retains the striped, flux-closed configuration and to a large extent insensitive to the applied voltage. Reversing the voltage polarity to +25 V immediately transforms the trenches near the center of the image (at the point of application of voltage) back to tetragonal-like morphology (fig 5(c)) and this trend continues on increasing the voltage to +100 V. We have to mention that this field-induced phase transition is also thickness dependent and for the 200 nm film the T-R/R-T transitions are observed only at very high voltages ( ± 50 volts or above, data not shown).

We now discuss the local ferroelectric switching properties of the nearly-tetragonal domains. As the discussions of the preceding paragraphs demonstrate, any voltage dependent measurements on tetragonal BFO samples are prone to structural phase transition between the tetragonal and the rhombohedral polymorphs. Therefore precautions are in order to properly assess the local FE switching properties of tetragonal BFO, and correct inferences can only be made if the local area being probed remains tetragonal throughout the voltage sweep. In the switching spectroscopy technique employed to measure the local ferroelectric loops a series of dc offset voltages are applied to the tip which is ramped up to a maximum voltage ($V_{max}$). Measurements performed at different $V_{max}$ are repeated 25 times to improve the signal-to-noise ratio and also to observe, if any, quasi-static time-dependent behavior. The applied dc voltage sweep is a triangular pattern starting from 0 to $+V_{max}$ to $-V_{max}$ to 0 with $V_{max}$ = 5, 10, 15, 20, 25V. Additionally, the topography is imaged after every hysteresis measurement to monitor any T-R switching. It is important to note that before completion of the amplitude cycle the voltage goes to $-V_{max}$. Therefore, any T-R switching event should be captured in the post-measurement topographic image as a rhombohedral trench. If not, we can conclude that the loop obtained is a measure of the pure tetragonal phase. Complete switching is observed at ±25 V as shown in Fig.6 (a) and (b) where we plot the averaged SS-PFM amplitude and phase loops for the 100 nm tetragonal BFO film. The symmetric butterfly loops of the amplitude scans and phase change from -1.5 radians (-90 degree) to almost +1.5 radians (+90 degrees) in Fig. 6(b) confirm a near complete polarization switching process in the tetragonal phase BFO. At tip voltages below 25V the switching process are partial with asymmetric amplitude loops and lower than 180 degree phase change. Quite noteworthy is the gradual phase change of the switching process again possibly due to the off-normal polarization direction in the film. We also performed additional SS-PFM measurements on the 80 nm BFO-R sample with bottom SRO electrodes, and the results are shown in figure 6(c) and (d). Extremely sharp polarization switching is observed in the phase loop with 180º phase change. Interestingly, the butterfly wings of the amplitude have an opposite sense compared to the BFO films deposited directly on LAO. The sharpness of the switching transition in BFO-R films, backed up by the PFM images, clearly indicate that the polarization switching process is distinctly different between rhombohedral and the nearly-tetragonal BFO films. It also needs to be emphasized that we have probed only the near-surface tetragonal domains. A more complicated switching involving intermediate states is possible if the entire bulk of the film is involved, given the distorted nature of tetragonal domains plus the mixed nature of the films. Phase field simulations such as performed on tetragonal BTO[32] and further experimental work on this nearly tetragonal phase are necessary to clarify this issue relating to the polarization reversal mechanism and its absolute value.

In summary, we have investigated thoroughly and demonstrate unambiguously that the nearly-tetragonal $BiFeO_3$ films grown on LAO substrates have a number of interesting ferroelectric and ferroelastic properties, true to its multiferroicity. Polarized Raman spectra show that the nearly tetragonal films have monoclinic (*Cc*) symmetry and not purely tetragonal (*P4mm*), and the rhombohedral phase becomes the dominant phase with increasing thickness. Through local switching-spectroscopy PFM measurements we show complete ferroelectric switching in tetragonal BFO, while vertical and lateral PFM measurements show that the ferroelectric domains of nearly-tetragonal BFO have a strong in-plane orientation, forming a striped, flux-closed configuration. Strong ferroelastic properties are demonstrated through electric-field driven phase transition between the tetragonal and rhombohedral polymorph.

The work at the University of Alabama was supported by ONR (Grant No. N00014-09-0119). A portion of this research was conducted at the Center for Nanophase Materials Sciences, which is sponsored at Oak Ridge National Laboratory by the Division of Scientific User Facilities, U.S. Department of Energy. The work of M.I. was supported by the State of Texas through the Texas Center for Superconductivity at the University of Houston.

## References

1. Wang, J.; Neaton, J. B.; Zheng, H.; Nagarajan, V.; Ogale, S. B.; Liu, B.; Viehland, D.; Vaithyanathan, V.; Schlom, D. G.; Waghmare, U. V.; Spaldin, N. A.; Rabe, K. M.; Wuttig, M.; Ramesh, R., Epitaxial $BiFeO_3$ Multiferroic Thin Film Heterostructures. *Science* **2003,** *299* (5613), 1719-1722.

2. Spaldin, N. A.; Fiebig, M., MATERIALS SCIENCE: The Renaissance of Magnetoelectric Multiferroics. *Science* **2005,** *309* (5733), 391-392.

3. Choi, T.; Lee, S.; Choi, Y. J.; Kiryukhin, V.; Cheong, S.-W., Switchable Ferroelectric Diode and Photovoltaic Effect in $BiFeO_3$. *Science* **2009,** *324* (5923), 63-66.

4. Zheng, H.; Wang, J.; Lofland, S. E.; Ma, Z.; Mohaddes-Ardabili, L.; Zhao, T.; Salamanca-Riba, L.; Shinde, S. R.; Ogale, S. B.; Bai, F.; Viehland, D.; Jia, Y.; Schlom, D. G.; Wuttig, M.; Roytburd, A.; Ramesh, R., Multiferroic $BaTiO_3$-$CoFe_2O_4$ Nanostructures. *Science* **2004,** *303* (5658), 661-663.

5. Bibes, M.; Barthelemy, A., Multiferroics: Towards a magnetoelectric memory. *Nat Mater* **2008,** *7* (6), 425-426.

6. Tsymbal, E. Y.; Kohlstedt, H., APPLIED PHYSICS: Tunneling Across a Ferroelectric. *Science* **2006,** *313* (5784), 181-183.

7. Velev, J. P.; Duan, C.-G.; Burton, J. D.; Smogunov, A.; Niranjan, M. K.; Tosatti, E.; Jaswal, S. S.; Tsymbal, E. Y., Magnetic Tunnel Junctions with Ferroelectric Barriers: Prediction of Four Resistance States from First Principles. *Nano Letters* **2008,** *9* (1), 427-432.

8. Gruverman, A.; Wu, D.; Lu, H.; Wang, Y.; Jang, H. W.; Folkman, C. M.; Zhuravlev, M. Y.; Felker, D.; Rzchowski, M.; Eom, C. B.; Tsymbal, E. Y., Tunneling Electroresistance Effect in Ferroelectric Tunnel Junctions at the Nanoscale. *Nano Letters* **2009,** *9* (10), 3539-3543.

9. Zhuravlev, M. Y.; Wang, Y.; Maekawa, S.; Tsymbal, E. Y., Tunneling electroresistance in ferroelectric tunnel junctions with a composite barrier. *Applied Physics Letters* **2009,** *95* (5), 052902-3.

10. Garcia, V.; Fusil, S.; Bouzehouane, K.; Enouz-Vedrenne, S.; Mathur, N. D.; Barthelemy, A.; Bibes, M., Giant tunnel electroresistance for non-destructive readout of ferroelectric states. *Nature* **2009,** *460* (7251), 81-84.

11. Maksymovych, P.; Jesse, S.; Yu, P.; Ramesh, R.; Baddorf, A. P.; Kalinin, S. V., Polarization Control of Electron Tunneling into Ferroelectric Surfaces. *Science* **2009,** *324* (5933), 1421-1425.

12. Streiffer, S. K.; Parker, C. B.; Romanov, A. E.; Lefevre, M. J.; Zhao, L.; Speck, J. S.; Pompe, W.; Foster, C. M.; Bai, G. R., Domain patterns in epitaxial rhombohedral ferroelectric films. I. Geometry and experiments. *Journal of Applied Physics* **1998,** *83* (5), 2742-2753.

13. Chu, Y.-H.; Cruz, M. P.; Yang, C.-H.; Martin, L. W.; Yang, P.-L.; Zhang, J.-X.; Lee, K.; Yu, P.; Chen, L.-Q.; Ramesh, R., Domain Control in Multiferroic $BiFeO_3$ through Substrate Vicinality. *Advanced Materials* **2007,** *19* (18), 2662-2666.

14. Jang, H. W.; Ortiz, D.; Baek, S.-H.; Folkman, C. M.; Das, R. R.; Shafer, P.; Chen, Y.; Nelson, C. T.; Pan, X.; Ramesh, R.; Eom, C.-B., Domain Engineering for Enhanced Ferroelectric Properties of Epitaxial (001) $BiFeO_3$ Thin Films. *Advanced Materials* **2009,** *21* (7), 817-823.

15. Vilas Shelke, D. M., Gopalan Srinivasan*,* Stephen Jesse*,* Sergei Kalinin*,* Arthur Baddorf *, and* Arunava Gupta, Reduced Coercive Field in $BiFeO_3$ Thin Films through Domain Engineering
*Submitted to Advanced Materials* **2010**.

16. Zeches, R. J.; Rossell, M. D.; Zhang, J. X.; Hatt, A. J.; He, Q.; Yang, C.-H.; Kumar, A.; Wang, C. H.; Melville, A.; Adamo, C.; Sheng, G.; Chu, Y.-H.; Ihlefeld, J. F.; Erni, R.; Ederer, C.; Gopalan, V.; Chen, L. Q.; Schlom, D. G.; Spaldin, N. A.; Martin, L. W.; Ramesh, R., A Strain-Driven Morphotropic Phase Boundary in $BiFeO_3$. *Science* **2009,** *326* (5955), 977-980.

17. Béa, H.; Dupé, B.; Fusil, S.; Mattana, R.; Jacquet, E.; Warot-Fonrose, B.; Wilhelm, F.; Rogalev, A.; Petit, S.; Cros, V.; Anane, A.; Petroff, F.; Bouzehouane, K.; Geneste, G.; Dkhil, B.; Lisenkov, S.; Ponomareva, I.; Bellaiche, L.; Bibes, M.; Barthélémy, A., Evidence for Room-Temperature Multiferroicity in a Compound with a Giant Axial Ratio. *Physical Review Letters* **2009,** *102* (21), 217603.

18. Ricinschi, D.; Yun, K.-Y.; Okuyama, M., A mechanism for the 150 $\mu C\ cm^{-2}$ polarization of $BiFeO_3$ films based on first-principles calculations and new structural data. *Journal of Physics: Condensed Matter* **2006,** *18* (6), L97-L105.

19. Ederer, C.; Spaldin, N. A., Effect of Epitaxial Strain on the Spontaneous Polarization of Thin Film Ferroelectrics. *Physical Review Letters* **2005,** *95* (25), 257601.

20. Ravindran, P.; Vidya, R.; Kjekshus, A.; Fjellvåg, H.; Eriksson, O., Theoretical investigation of magnetoelectric behavior in $BiFeO_3$. *Physical Review B* **2006,** *74* (22), 224412.

21. Guo, R.; Cross, L. E.; Park, S. E.; Noheda, B.; Cox, D. E.; Shirane, G., Origin of the High Piezoelectric Response in $PbZr_{1-x}Ti_xO_3$. *Physical Review Letters* **2000,** *84* (23), 5423.

22. Ahart, M.; Somayazulu, M.; Cohen, R. E.; Ganesh, P.; Dera, P.; Mao, H.-k.; Hemley, R. J.; Ren, Y.; Liermann, P.; Wu, Z., Origin of morphotropic phase boundaries in ferroelectrics. *Nature* **2008,** *451* (7178), 545-548.

23. Park, S.-E.; Shrout, T. R., Ultrahigh strain and piezoelectric behavior in relaxor based ferroelectric single crystals. *Journal of Applied Physics* **1997,** *82* (4), 1804-1811.

24. Shelke, V.; Harshan, V. N.; Kotru, S.; Gupta, A., Effect of kinetic growth parameters on leakage current and ferroelectric behavior of $BiFeO_3$ thin films. *Journal of Applied Physics* **2009,** *106* (10), 104114-7.

25. Hatt, A. J.; Spaldin, N. A.; Ederer, C., Strain-induced isosymmetric phase transition in $BiFeO_3$. *Physical Review B 81* (5), 054109.

26. Shelke, V.; Srinivasan, G.; Gupta, A., Ferroelectric properties of $BiFeO_3$ thin films deposited on substrates with large lattice mismatch. *physica status solidi (RRL) - Rapid Research Letters 4*, 79-81.

27. Iliev, M. N. A., M. V.; Mazumdar, D.; Shelke.V.; Gupta, A., *unpublished data* **2010**.

28. Jesse, S.; Lee, H. N.; Kalinin, S. V., Quantitative mapping of switching behavior in piezoresponse force microscopy. *Review of Scientific Instruments* **2006,** *77* (7), 073702-10.

29. Jesse, S.; Rodriguez, B. J.; Choudhury, S.; Baddorf, A. P.; Vrejoiu, I.; Hesse, D.; Alexe, M.; Eliseev, E. A.; Morozovska, A. N.; Zhang, J.; Chen, L.-Q.; Kalinin, S. V., Direct imaging of the spatial and energy distribution of nucleation centres in ferroelectric materials. *Nat Mater* **2008,** *7* (3), 209-215.

30. Catalan, G.; Béa, H.; Fusil, S.; Bibes, M.; Paruch, P.; Barthélémy, A.; Scott, J. F., Fractal Dimension and Size Scaling of Domains in Thin Films of Multiferroic $BiFeO_3$. *Physical Review Letters* **2008,** *100* (2), 027602.

31. Dipanjan Mazumdar, V. S., Stephen Jesse*,* Sergei Kalinin*,* Arthur Baddorf *, and* Arunava Gupta,, *unpublished data* **2009**.

32. Li, Y. L.; Chen, L. Q., Temperature-strain phase diagram for $BaTiO_3$ thin films. *Applied Physics Letters* **2006,** *88* (7), 072905-3.

## Figure Captions

Figure 1. (color online) (a) 2θ-θ scans in the 35-50º range for 70, 100, 200 nm BFO films deposited directly on LAO substrates and an 80 nm film on 40 nm SRO buffer layer. (b) Complete $\phi$-scan of the (101) reflection for a 70 nm BFO film on LAO substrate. Inset shows the sub-features of one of the $\phi$ peaks showing a three-variant structure. (c) Reciprocal space map (RSM) for a 70 nm BFO film on LAO substrate near the (103) reflection of the tetragonal phase. RSM also show a three variant structure. (d) Schematic diagram of the tetragonal and pseudo-cubic rhombohedral unit cell scaled according to the tetragonality 'c/a' ratio.

Figure 2. (color online) Polarized Raman spectra of (a) 70 nm BFO/LAO, (b) 100 nm BFO/LAO, (c) 200 nm BFO/LAO, and (d) 80 nm BFO/40 nm SRO/LAO thin films. *x*, *y*, *x'* and *y'* are parallel, respectively, to the $[100]_c$, $[010]_c$, $[110]_c$, and $[-110]_c$ quasi-cubic directions of the LAO substrate.

Figure 3. (color online) Topography (i), out-of-plane PFM amplitude (ii), and PFM phase (iii) micrographs of (a) BFO(80 nm)/SRO(40 nm)/LAO (b) BFO(100 nm)/LAO (c) BFO(200 nm)/LAO. SRO buffered BFO show mosaic domain feature where as the tetragonal domains show predominantly in-plane, flux-closed stripe domains (see supplementary figures S1-S3).

Figure 4. (color online) (a) Topographic image of the 100 nm BFO film in the as-deposited condition (b) Topography after the application of a -20V 1 sec pulse at the center (c) Topography after +20V (d) after -20V. The reversibility of the transition is clearly demonstrated. The trenches resemble R-phase trenches of the as-deposited state. (e) Energy landscape diagram showing the difference in the energy between BFO-T and BFO-R bulk phases (left) and in the nearly-tetragonal thin films (right).

Figure 5. Topography and PFM amplitude/ phase data of a 1μm by 1μm area of the 100 nm BFO sample after the application of (a) -25V (b) -75V (c) +25V (d) +75V at the center. Thick black lines in figure b(iii) are a guide to the striped, flux-closed configuration of the tetragonal BFO film.

Figure 6. Local switching-spectroscopy PFM amplitude and phase data of 100 nm nearly-tetragonal BFO film (a and b), and SRO-buffered rhombohedral BFO films (c and d). Symmetric butterfly loops and phase change value of 180 degrees (~ 3 pi radians) imply complete polarization switching in both films.

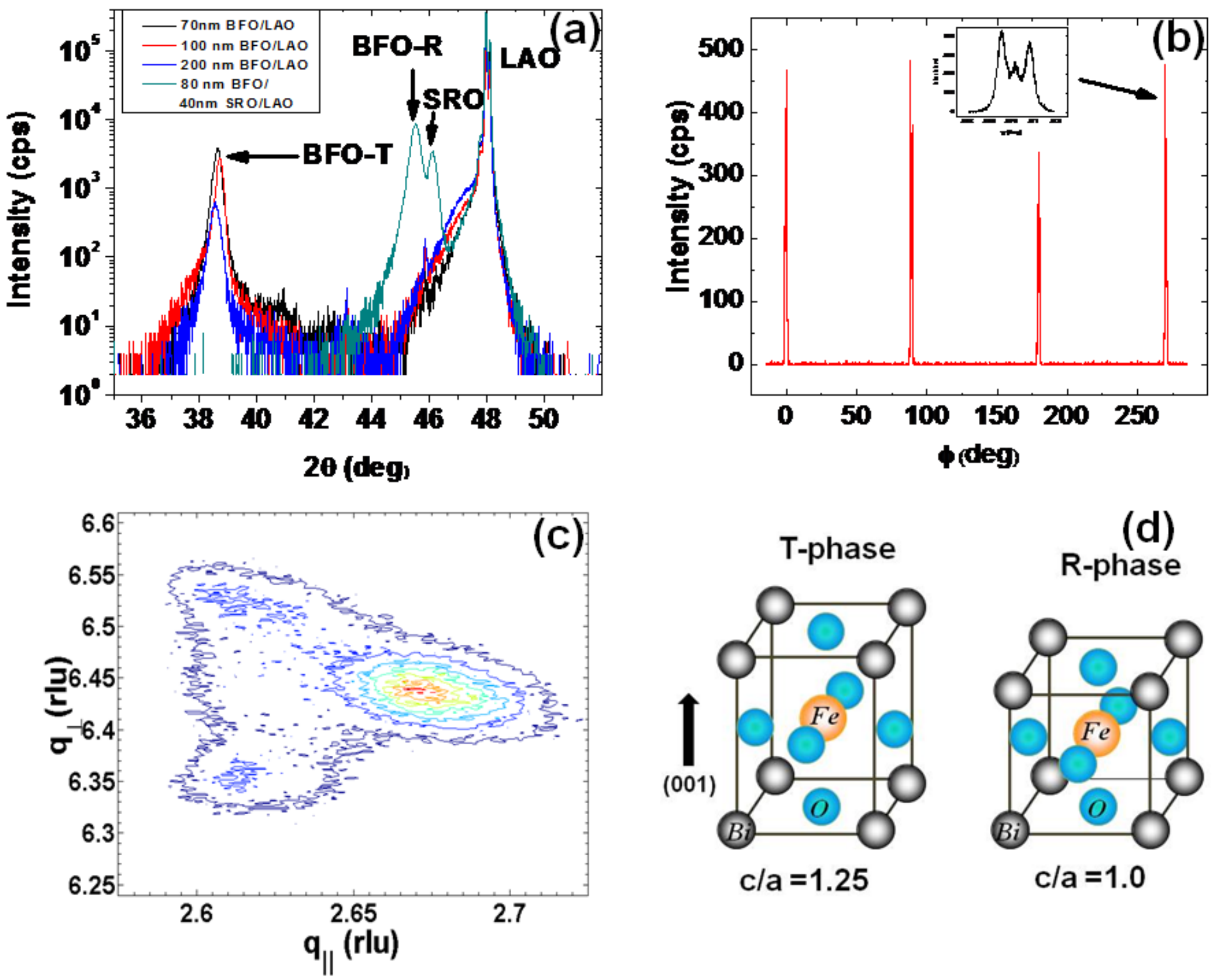
70nm BFO/LAO
100 nm BFO/LAO
200 nm BFO/LAO
80 nm BFO/
40nm SRO/LAO
(a)
BFO-R
SRO
LAO
BFO-T
Intensity (cps)
2θ (deg)
(b)
Intensity (cps)
φ (deg)
(c)
q⊥ (rlu)
q|| (rlu)
(d)
T-phase
R-phase
(001)
Fe
O
Bi
c/a =1.25
c/a =1.0

Figure 1

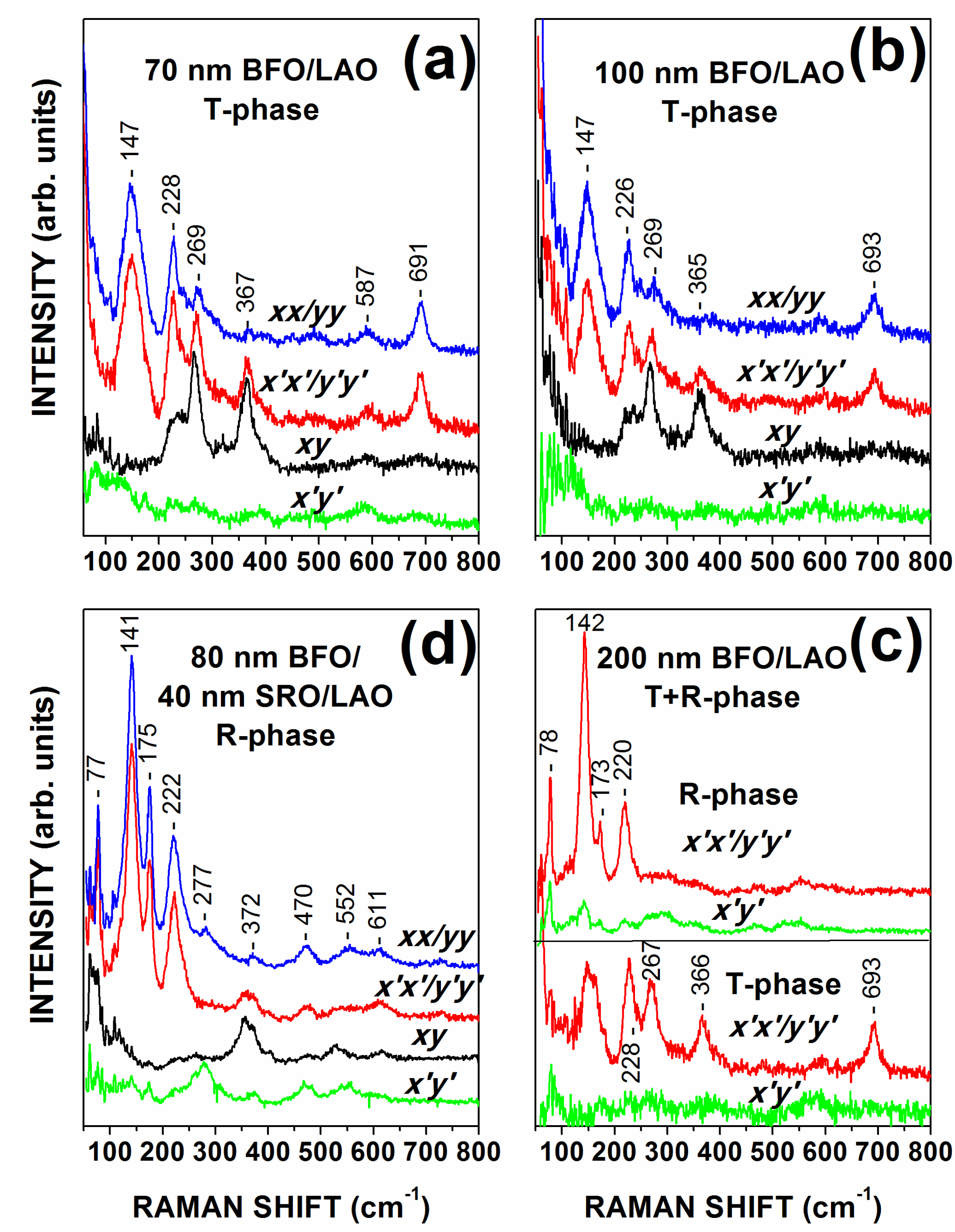

70 nm BFO/LAO
T-phase
(a)
147
228
269
367
587
691
xx/yy
x'x'/y'y'
xy
x'y'
100 nm BFO/LAO
T-phase
(b)
147
226
269
365
693
xx/yy
x'x'/y'y'
xy
x'y'
80 nm BFO/
40 nm SRO/LAO
R-phase
(d)
77
141
175
222
277
372
470
552
611
xx/yy
x'x'/y'y'
xy
x'y'
200 nm BFO/LAO
T+R-phase
(c)
142
78
173
220
R-phase
x'x'/y'y'
x'y'
267
366
693
228
T-phase
x'x'/y'y'
x'y'
INTENSITY (arb. units)
RAMAN SHIFT (cm$^{-1}$)
100 200 300 400 500 600 700 800

Figure 2

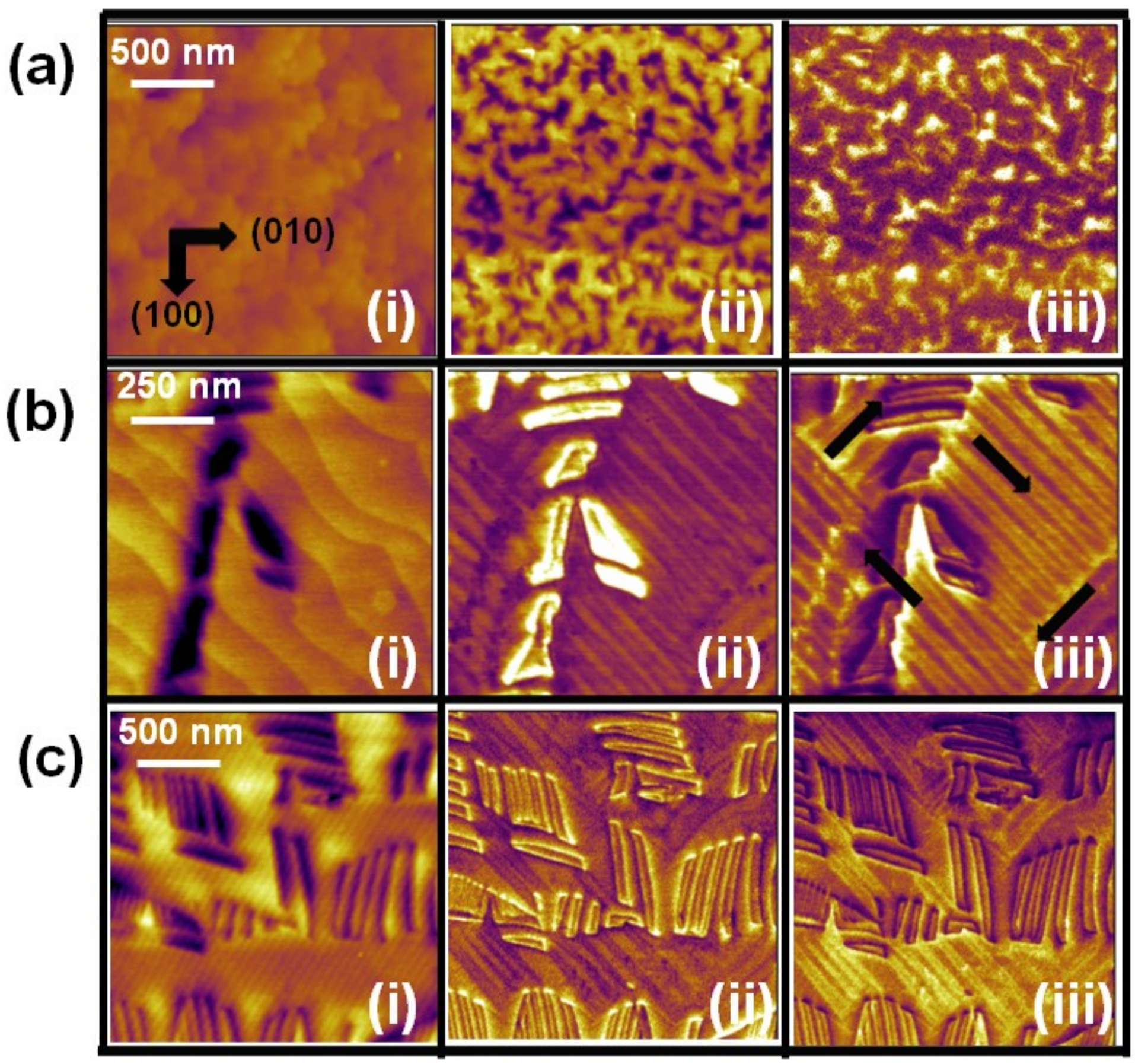

Figure 3

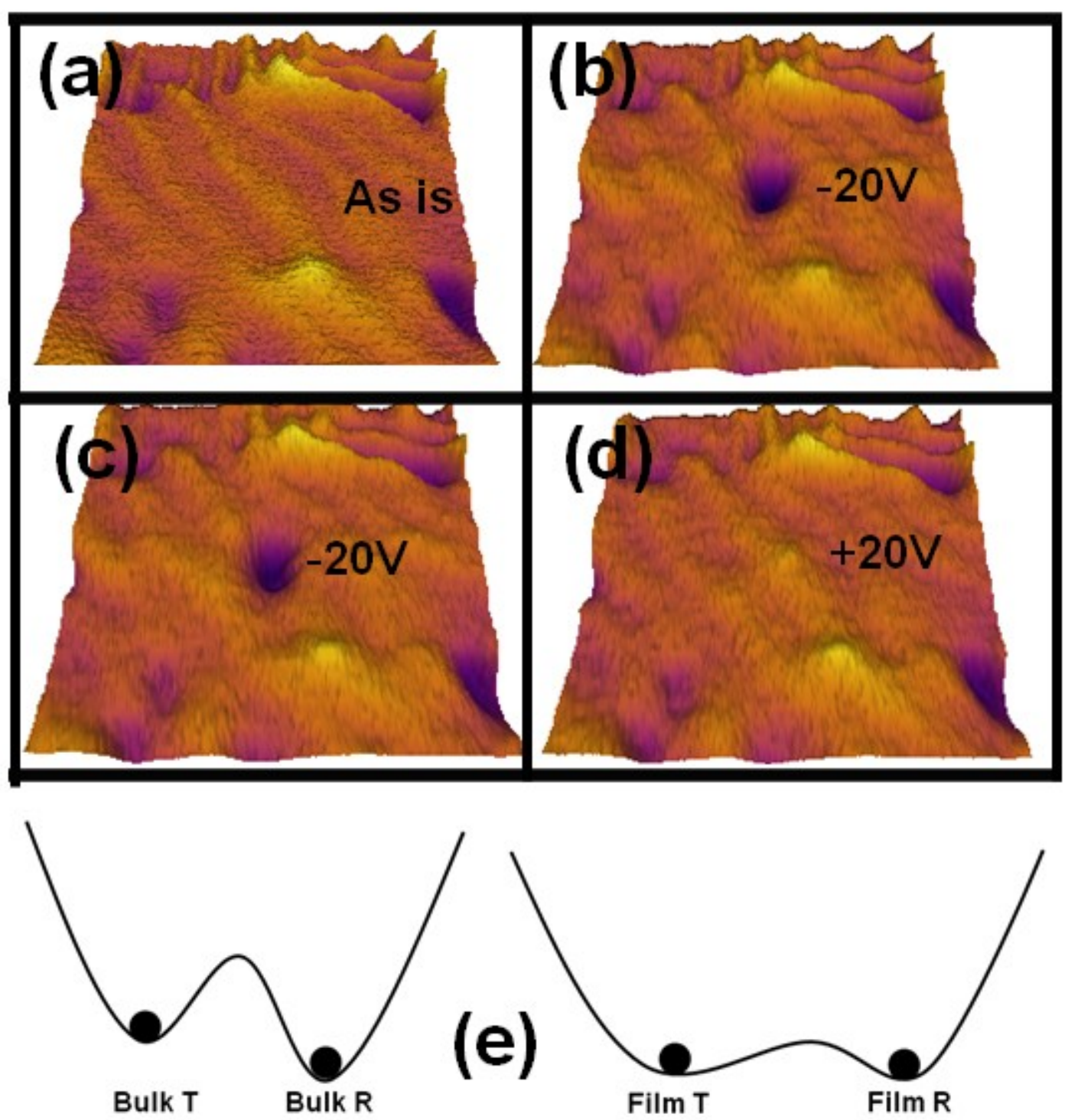

Figure 4

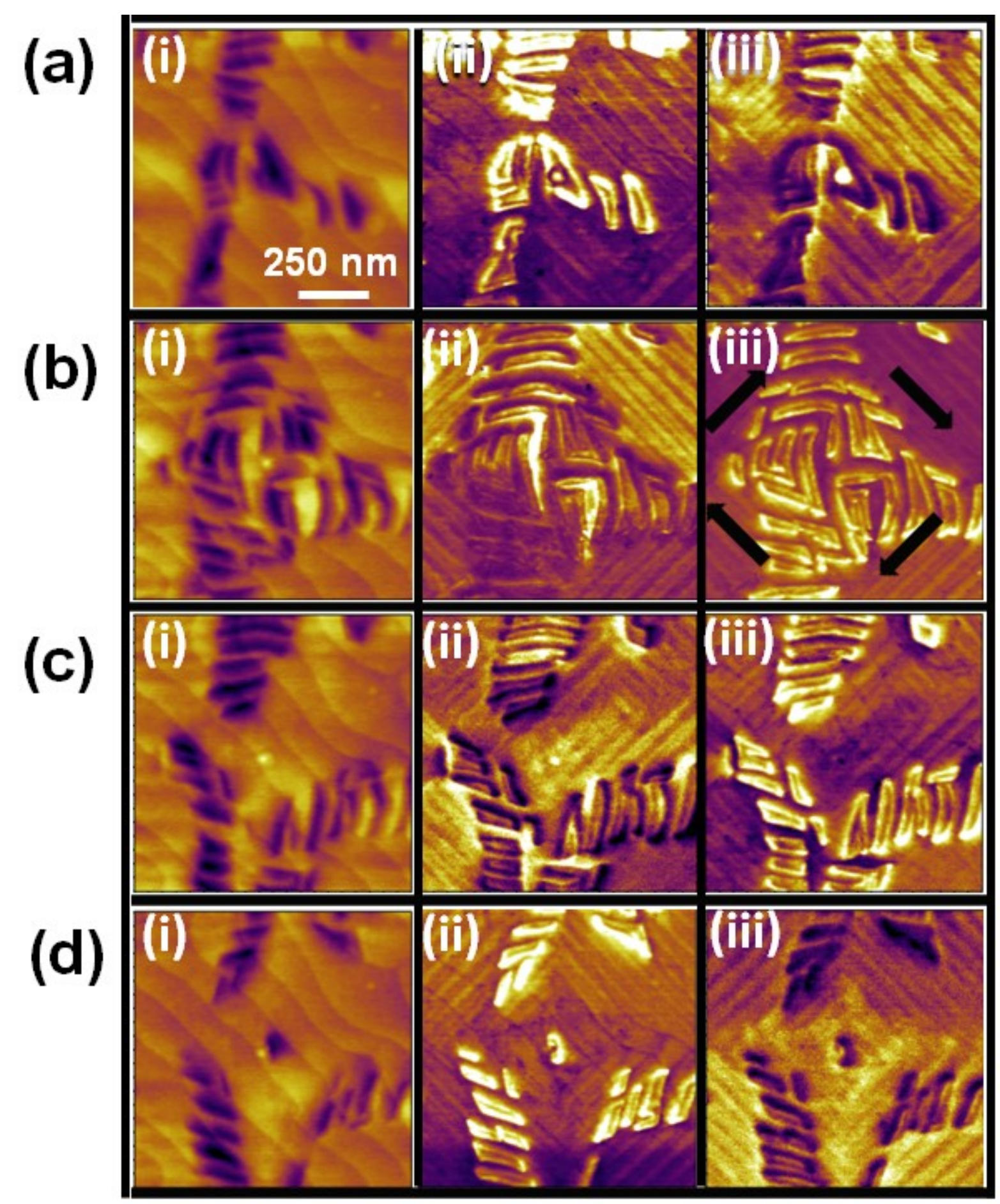

Figure 5

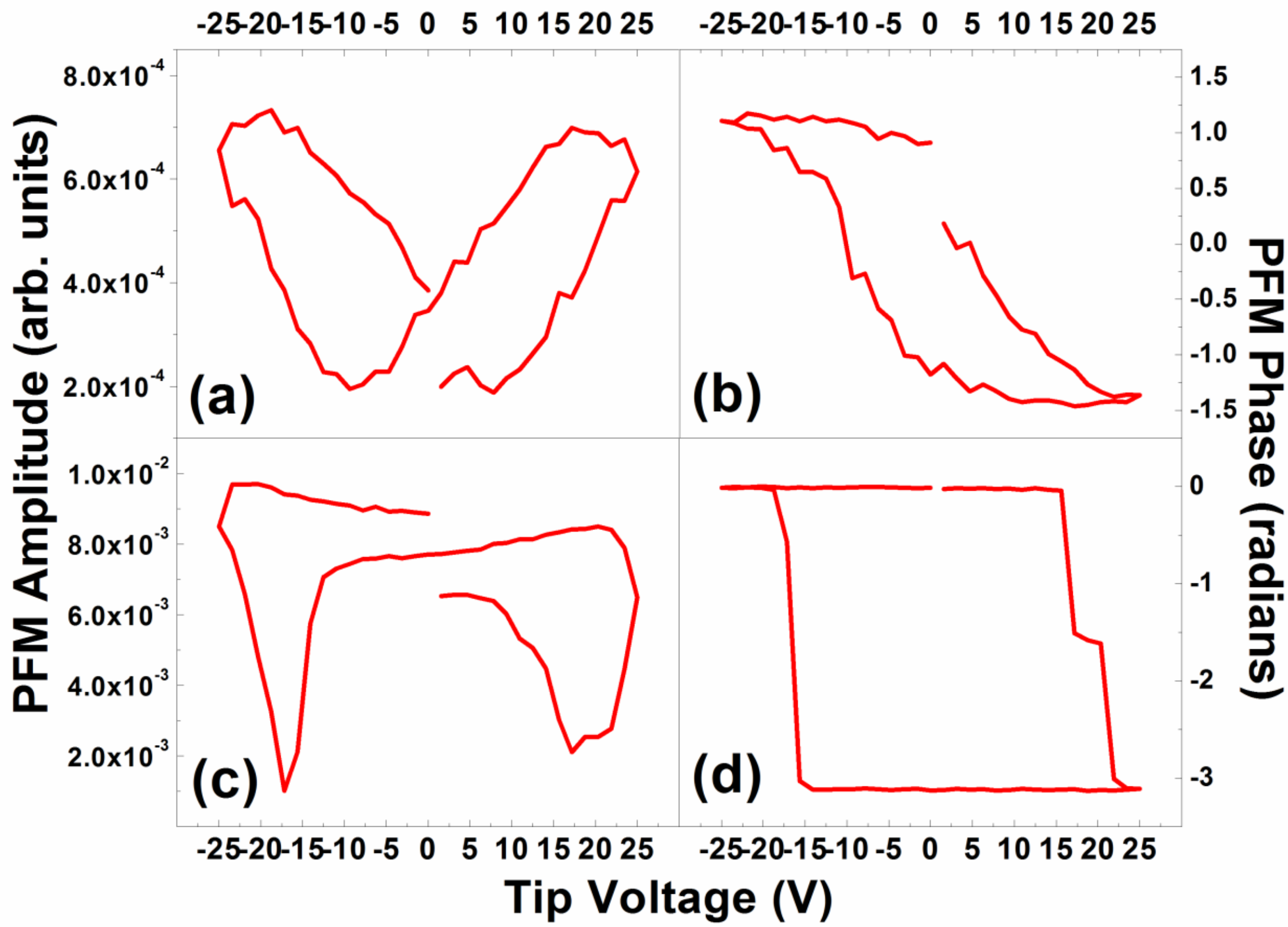
-25-20-15-10 -5 0 5 10 15 20 25
8.0x10^-4
6.0x10^-4
4.0x10^-4
2.0x10^-4
1.0x10^-2
8.0x10^-3
6.0x10^-3
4.0x10^-3
2.0x10^-3
PFM Amplitude (arb. units)
1.5
1.0
0.5
0.0
-0.5
-1.0
-1.5
0
-1
-2
-3
PFM Phase (radians)
(a)
(b)
(c)
(d)
Tip Voltage (V)

Figure 6